\documentstyle[emulateapj,graphicx]{article}
\def\gsim{\;\rlap{\lower 2.5pt
 \hbox{$\sim$}}\raise 1.5pt\hbox{$>$}\;}
\def\lsim{\;\rlap{\lower 2.5pt
   \hbox{$\sim$}}\raise 1.5pt\hbox{$<$}\;}

\newcommand{\beq}{\begin{equation}}
\newcommand{\eeq}{\end{equation}}
\def\myputfigure#1#2#3#4#5%
{\hskip0.03\textwidth\vskip#5pt
\makebox[0pt]{\hskip#2in
\includegraphics[width=#3\textwidth]{#1}}\vskip#4pt\hfill}


\lefthead{Haiman}
\righthead{Growth of Supermassive Black Holes with Gravitational Recoil}

\begin{document}
\title{Constraints from Gravitational Recoil on the Growth of \\ Supermassive Black Holes at High Redshift}
\author{Zolt\'an Haiman\altaffilmark{1}}
\affil{Department of Astronomy, Columbia University, 550 West 120th Street, New York, NY 10027\\
zoltan@astro.columbia.edu}

\vspace{\baselineskip}
\submitted{To be submitted to ApJ Letters}
\begin{abstract}
Recent studies have shown that during their coalescence, binary
supermassive black holes (SMBHs) experience a gravitational recoil
with velocities of $100~{\rm km~s^{-1}}\lsim v_{\rm kick} \lsim
600~{\rm km~s^{-1}}$.  These velocities exceed the escape velocity
$v_{\rm esc}$ from typical dark matter (DM) halos at high--redshift
($z\gsim 6$), and therefore put constraints on scenarios in which
early SMBHs grow at the centers of DM halos.  Here we quantify these
constraints for the most distant known SMBHs, with inferred masses in
excess of $10^9{\rm M_\odot}$, powering the bright quasars discovered
in the Sloan Digital Sky Survey at $z>6$.  We assume that these SMBHs
grew via a combination of accretion and mergers between pre--existing
seed BHs in individual progenitor halos, and that mergers between
progenitors with $v_{\rm esc}< v_{\rm kick}$ disrupt the BH growth
process.  Our results suggest that under these assumptions, the $z\sim
6$ SMBHs had a phase during which gained mass significantly more
rapidly than under an Eddington--limited exponential growth rate.
\end{abstract}

\section{Introduction}

The gravitational waves (GWs) emitted during the final stages of the
coalescence of two merging black holes (BHs) carry linear momentum,
implying that the center of mass of the system experiences a recoil
(Bonnor \& Rotenberg 1961; Peres 1962).  The resulting recoil velocity
had been known to be large, with values estimated to be of order
$v_{\rm kick} \sim 1000~{\rm km~s^{-1}}$ (Fitchett 1983). Favata et
al. (2004) has recently revisited this problem and computed recoil
velocities, treating the spin and orbital dynamics of the merging BHs,
as well as the generation of GWs in the strong gravity regime. They
have found the range of possible recoil velocities to be $100~{\rm
km~s^{-1}}\lsim v_{\rm kick} \lsim 600~{\rm km~s^{-1}}$, with the
exact value depending on the mass ratio of the merging BHs, their
spin, and orbital parameters.  These velocities are large compared to
the escape velocities of dwarf galaxies, and of the typical dark
matter halos that existed at the early epochs of galaxy formation
($z\gsim 6$).  In hierarchical cosmogonies, the SMBHs that are known
to exist in the local universe grew via a combination of accretion and
mergers between holes residing in individual DM halos.  Merritt et
al. (2004) and Madau \& Quataert (2004) recently considered several
consequences of a large recoil that removes a BH from the center of
its host galaxy.  In particular, Merritt et al. (2004) pointed out
that the ejection of SMBHs from the shallow potentials of DM halos at
high redshift implies a maximum redshift at which the progenitors of
present--day SMBHs could have started merging (and sticking) with each
other.

In this {\em Letter}, we consider the growth history of SMBHs that are
in place at $z>6$, and are thought to power the bright quasars
recently discovered (Fan et al. 2000; 2001; 2003) in the Sloan Digital
Sky Survey (SDSS).  As discussed in Haiman \& Loeb (2001; hereafter
HL01), relatively little time is available for the growth of these few
$\times10^9~{\rm M_\odot}$ SMBHs prior to $z\sim 6$, and their seed
BHs must be present as early as $z\sim 10$.  A model in which stellar
seed BHs appear in small progenitor DM halos is consistent with the
presence of a $\sim 4\times10^9~{\rm M_\odot}$ SMBH at $z\sim 10$,
provided that each seed BH can grow at least at the Eddington--limited
exponential rate, and that the progenitor halos can form seed BHs
sufficiently eary on.

The ejection of a merger--product SMBH from its host halo severely
limits the ability of massive SMBHs to grow at $z>6$, by disrupting
the early stages of growth.  In this {\em Letter}, as an example, we
model the growth of the SMBH powering the most distant SDSS quasar,
SDSS 1054+1024 at redshift $z=6.43$, with an inferred BH mass of $\sim
4\times 10^9~{\rm M_\odot}$. Under the assumption that progenitor
holes are ejected from DM halos with velocity dispersions
$\sigma<v_{\rm kick}/2$, and do not contribute to the final BH mass,
we find that typical recoil velocities must either be below the {\it
minimum} value $v_{\rm kick}= 100~{\rm km~s^{-1}}$ found by Favata et
al., or else this SMBH must have had a phase during which it gained
mass significantly more rapidly than the Eddington--limited
exponential growth rate would imply.

The rest of this {\it Letter} is organized as follows.
In \S~\ref{sec:quasar}, we discuss the inferred values of the relevant
parameters (halo and BH mass) of SDSS 1054+1024.
In \S~\ref{sec:model}, we describe the method we use model the growth
of the SMBH by accretion and mergers.
In \S~\ref{sec:results}, we present our main result, showing that
excluding seed BHs from low--mass progenitor halos necessitates a 
faster--than Eddington growth rate.
In \S~\ref{sec:discuss}, we discuss various uncertainties about our
results.
In \S~\ref{sec:conclude}, we summarize the implications of this work
and offer our conclusions.\footnote{Throughout this paper, we adopt
the background cosmological parameters for a flat universe as measured
by the {\it WMAP} experiment, $\Omega_m=0.29$,
$\Omega_{\Lambda}=0.71$, $\Omega_b=0.047$, $h=0.72$, with a power
spectrum normalization $\sigma_{8h^{-1}}=0.9$ and slope $n=0.99$
(Spergel et al. 2003).}

\section{The Black Hole and Dark Matter of Halo of the $z=6.43$ Quasar SDSS 1054+1024}
\label{sec:quasar}

The starting point for the constraints we derive below is simply the
existence of a BH of mass $M_{\rm bh}$ at redshift $z$, residing
within a dark halo of mass $M_{\rm halo}$.  While the masses of SMBHs
at the centers of nearby galaxies can be directly estimated, the mass
of the SMBH powering the $z=6.43$ quasar SDSS 1054+1024 is inferred
indirectly from its observed luminosity.  Under the assumption that
the quasar emits a fraction $\eta$ of the Eddington luminosity, and
using the template spectrum of Elvis et al. (1994) to make a
bolometric correction, in the case of SDSS 1044-0125, we find $M_{\rm
bh}=4.6 \times 10^9\eta^{-1}{\rm M_\odot}$.  Given a sufficient
fueling rate, bright quasars would naturally shine at their limiting
luminosity, and we expect $\eta=1$.  There is no obvious signs of
beaming or lensing in the spectrum of this quasar (Willott et
al. 2003).  Indeed, the large observed size ($\sim 6$ comoving Mpc;
Mesinger \& Haiman 2004) of its Str\"omgren sphere makes it unlikely
that the apparent flux of this quasar was significantly boosted by
either lensing or beaming (Haiman \& Cen 2002).  In most conventional
accretion models, $\eta\leq 1$, and the fiducial value of $M_{\rm
bh}=4.6 \times 10^9 {\rm M_\odot}$ would be a lower limit to the
actual BH mass.  However, there are models with $\eta>1$; for example,
in the recent ``photon bubble'' model of Begelman (2002), $\eta$ can
be as high as $\eta\sim 10$, reducing the inferred BH mass.

We next require the mass of the halo in which the quasar SDSS
1054+1024 resides.  As in the case of the BH mass, the halo mass, or
the velocity dispersion, can be directly measured for some nearby
AGNs, but for distant quasars, we have to rely on indirect estimates.
As described in HL01, $M_{\rm halo}$ can be estimated based on the
abundance of dark matter halos. In the case of the SDSS quasar, one
bright quasar was found within a $\approx 2000$ deg$^2$ survey area.
In order to match this abundance, we find the halo mass has to be
$M_{\rm halo}\approx 8.5\times10^{12}~{\rm M_\odot}$, with a
corresponding velocity dispersion of $420~{\rm km~s^{-1}}$. This
result is weakly sensitive to the assumed duty cycle of quasar
activity.  We here assume $t_{\rm Q}=4\times 10^7
(\epsilon/0.1)\eta^{-1}~{\rm yr}$, where $\epsilon$ is the usual
radiative efficiency of accretion.  We use the standard
Press--Schechter mass function (but find that our results would change
little if we had instead adopted the recent numerical mass function in
Jenkins et al. 2002, or the improved semi--analytical mass function of
Sheth \& Tormen 1999). Further details of the BH and halo mass
determination are given in HL01.

A relation between BH mass and halo circular velocity was recently
determined in a sample of nearby galaxies (Ferrarese 2002). While this
local relation does not necessarily hold at higher redshifts, it is
interesting to note that it is in good agreement with the BH mass and
(halo) velocity dispersion our procedure yields for SDSS 1054+1024.
Finally, the inferred halo mass is consistent with the value of
$M_{\rm halo}\sim 10^{13}~{\rm M_\odot}$ derived from the spectral
signatures of cosmic infall for this source (Barkana \& Loeb 2003).

\section{Growth of Black Holes in Hierarchical Model}
\label{sec:model}

In order to model the growth of the SMBH in our adopted $\Lambda$CDM
cosmology, we rely on the merger history of dark matter halos in the
extended Press--Schechter (EPS) formalism (Press \& Schechter 1974;
Lacey \& Cole 1993).  We follow HL01, and compute the central BH mass
mirroring the assembly of its host halo. Given a parent halo of total
mass $M_{\rm halo}$ at redshift $z$, the EPS formalism specifies its
average merger history back in redshift.  Every branch of such a
merger tree represents a progenitor of the parent halo, whose mass is
continuously growing by accreting, and by merging with other
progenitors.

To keep our model simple, we assume that each progenitor of the parent
halo develops a seed BH of mass $M_{\rm seed}$ when the progenitor
grows above a critical size, corresponding to the velocity dispersion
$\sigma_{\rm min}=v_{\rm kick}/2$.  Here $v_{\rm kick}$ is a typical
recoil velocity for a coalescing SMBH binary.  Any seed BH that had
appeared further up along the branch in the merger tree (e.g. the
remnants of the first stars; Abel et al. 2002; Bromm et al. 2002) are
thus assumed to be ejected from the progenitor halo, and not to
contribute to the final SMBH mass at $z\sim 6$.  This is a reasonable
assumption, since in a typical ``merger tree'', each progenitor halo
had been continuously undergoing mergers with other progenitors.  The
expected recoil velocity depends on the mass--ratio of the merging BHs;
this dependence could be included in more detailed models of the
merger tree, such as those utilizing Monte-Carlo realizations
(Kauffmann \& Haehnelt 2000; Menou et al. 2001; Volonteri et al. 2003;
Islam et al. 2003). Once the progenitor halo has merged with several
others, its potential well will be deep enough so that the ejected
seed BH could fall back into this enlarged halo. However, the ejected
BH will likely have traveled far beyond the virial radius of the
enlarged halo: we find $t_{\rm Hub}(z)v_{\rm kick} / R_{\rm
vir}(z,\sigma)\gsim 10$ for $z\sim 10$ and $\sigma\sim 50~{\rm
km~s^{-1}}$.  Madau \& Quataert (2004) recently considered the
dynamics of a recoiling BH in the fixed potential of the host galaxy,
and showed that for a density distribution with a steep stellar cusp,
dynamical friction can cause the BH to return to the halo center
within $\sim 10^6$ yrs.  It is unlikely, however, that steep stellar
cusps exist in the earliest proto--halos that are thought to form a
single massive star at $z\sim 17-18$ (which, as we will see below, is
the birth redshift for the seeds responsible for the bulk of the final
SMBH mass).

We next assume that each individual seed BH subsequently grows
exponentially by accretion, $M_{\rm bh}(t)= \exp[\Delta t(z,
z_f)/t_{\rm acc}] M_{\rm seed}$, where $t_{\rm acc}=4\times 10^7
(\epsilon/0.1)\eta^{-1}~{\rm yr}$, as defined above, and $\Delta t$ is
the time elapsed between the formation of the seed BH at redshift
$z_f$ and a later redshift $z$.  We assume that eventually, by
redshift $z$, the smaller BHs in all progenitor halos coalesce
together to form a single SMBH at the center of the parent halo (as
long as the BH mergers are completed prior to redshift $z$, we do not
need to specify when they take place).  The mass of the resulting SMBH
in the parent halo at redshift $z$ is the sum of the individual BHs,
each of which has grown by a different amount:

\begin{equation}
M_{\rm bh}(z,M_{\rm halo}) = M_{\rm seed} \int_z^\infty dz^\prime 
\frac{dN_{\rm prog}}{dz^\prime} 
\exp\left[\frac{\Delta t(z,z^\prime)}{t_{\rm acc}}\right],
\label{eq:Mbh}
\end{equation}

where $N_{\rm prog}(z^\prime)$ is the number of seeded
progenitors at redshift $z^\prime > z$,

\begin{equation}
N_{\rm prog}(z^\prime)=\int_{M_{\rm min}}^{M_{\rm halo}} dM 
\frac{dP(z,z^\prime,M_{\rm halo},M)}{dM}.
\label{eq:Nprog}
\end{equation}

Here $dP(z,z^\prime,M_{\rm halo},M)$ is the number of progenitors of
mass $M$ at redshift $z^\prime$ of a halo whose mass at redshift $z$
is $M_{\rm halo}$ (Lacey \& Cole 1993, eq. 2.15), and $M_{\rm min}(z)$
is the mass of a halo whose velocity dispersion is $v_{\rm kick}/2$.

To summarize, our model for the assembly of BHs has five parameters.
Two of these, $M_{\rm halo}$ and $\eta$, describe the observed quasar
SDSS 1054+1024, and have relatively small uncertainties, as discussed
in the previous section.  The three parameters $M_{\rm seed}$,
$\epsilon$, and $v_{\rm kick}$ relate to our model for the growth of
the SMBH.  The fiducial values of these parameters are chosen as
follows.  The seed mass is $M_{\rm seed}=10~{\rm M_\odot}$, the
typical value for a stellar remnant BH.  VMOs (Carr, Bond \& Arnett
1984) can leave larger remnants, weighing up to $\sim 10^3~{\rm
M_\odot}$ (Heger et al. 2003).  The radiative efficiency is taken to
be $\epsilon=0.1$, based on the last stable orbit around a
non--rotating BH.  This value is consistent with a comparison of
quasar light to remnant BH masses in nearby galaxies (Yu \& Tremaine
2002; Aller \& Richstone 2002; Haiman, Ciotti \& Ostriker 2004).  A
maximally rotating Kerr BH would produce a larger value of
$\epsilon=0.42$.

For any given values of the above five parameters,
equation~\ref{eq:Mbh} can be used to compute $M_{\rm bh}=M_{\rm bh}(
M_{\rm halo},\eta, M_{\rm seed},\epsilon, v_{\rm kick})$. By requiring
the predicted BH mass to equal the value inferred from observations,
this relation can be inverted, and our model then yields a unique
prediction for $v_{\rm kick}$ as a function of the five parameters
$M_{\rm halo}$, $\eta$, $M_{\rm seed}$, $\epsilon$, and $M_{\rm bh}$.

\section{Results}
\label{sec:results}

In our fiducial model, we find numerically that the maximum recoil
velocity that allows the growth of the SMBH in the quasar SDSS
1054+1024 is $v_{\rm kick}=64~{\rm km~s^{-1}}$.  This value is
significantly below the lowest values predicted by Favata et
al. (2004) and Merritt et al. (2004). If actual recoil velocities are in
excess of $100~{\rm km~s^{-1}}$, this would be inconsistent with the
fiducial SMBH growth model presented here, and would require that some
of the seeds grow their mass faster than the assumed Eddington rate.

In order to illustrate the BH growth process in our model in somewhat
more detail, Figure~\ref{fig:sdss} shows the evolution of various
quantities for SDSS 1054+1024 (this figure is an updated version of
Figure 1 in HL01, which presented similar results for the earlier SDSS
quasar 1054+1000 at $z=5.8$).  In this figure, we have assumed $M_{\rm
seed}=10, \epsilon=0.1, \eta=1$, and $v_{\rm kick}=64~{\rm
km~s^{-1}}$.  With this combination, equations~\ref{eq:Mbh} and
\ref{eq:Nprog} yield the required BH mass of $M_{\rm bh}=4.6 \times
10^9~{\rm M_\odot}$ at $z=6.43$.  The top left panel in
Figure~\ref{fig:sdss} shows the number of progenitors of the parent
halo ($M_{\rm halo}\approx 8.5\times10^{12}~{\rm M_\odot}$) whose
velocity dispersion exceed $32~{\rm km~s^{-1}}$.  For reference, the
bottom left panel shows the corresponding minimum halo mass. Going
towards higher redshift, the number of progenitors increases, peaks at
$z\approx 11$, and then decreases again as the typical progenitors are
broken up into halos smaller than $32~{\rm km~s^{-1}}$.  The top right
panel shows the contribution of progenitors from each redshift to the
final black hole mass, and shows that the bulk of the BH mass is
contributed by seed holes from $17\lsim z\lsim 18$.  There are no new
seeds forming at $z\lsim 11$, and the peak redshift is considerably
higher than the peak at which most progenitors form.  This is simply
because the increased time available between $z=6.43$ and increasingly
higher redshifts $z$ (shown explicitly in the bottom right panel)
makes the contribution from the first $\sim 20$ progenitors, forming
at $z\sim 18$, dominant.

\vspace{\baselineskip}
\myputfigure{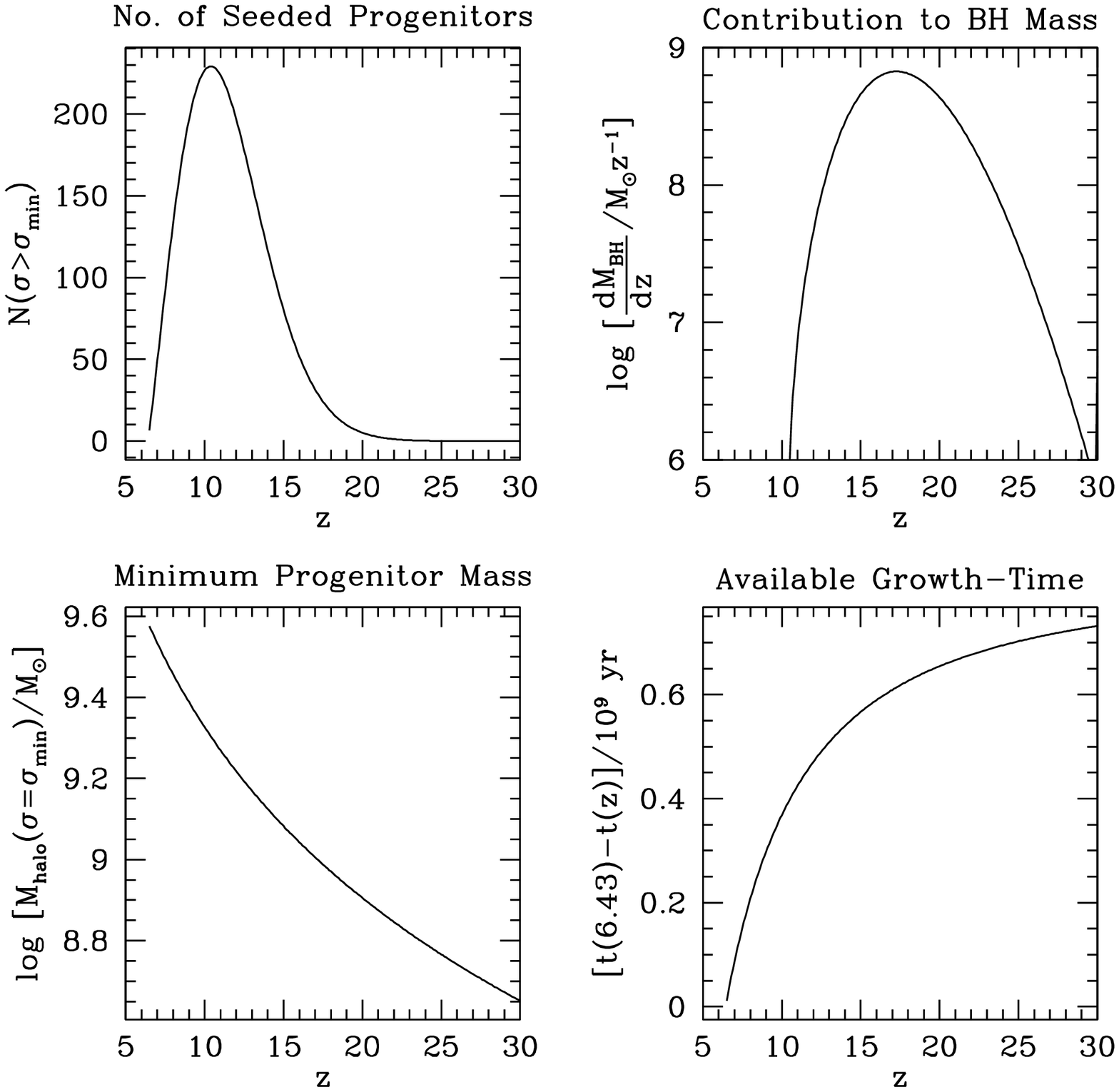}{3.2}{0.5}{-10}{-10} \figcaption{The
assembly history of the black hole in the $z=6.43$ quasar SDSS
1054+1024.  The inferred BH mass is $4.6 \times 10^9~{\rm M_\odot}$,
and the host halo mass is $8.5\times10^{12}~{\rm M_\odot}$. The four
panels show, clockwise, the number of seeded progenitors (i.e. those
with velocity dispersion above $32~{\rm km~s^{-1}}$); the contribution
of progenitors at different redshifts to the final BH mass at
$z=6.43$; the time available for the exponential growth of a seed
between $z$ and redshift of $6.43$; and the halo mass corresponding to
$32~{\rm km~s^{-1}}$ at each redshift.
\label{fig:sdss}}

 
\section{Discussion}
\label{sec:discuss}

We have found above that, in order to grow an SMBH as massive as
$4.6\times10^9~{\rm M_\odot}$ in our fiducial model, we would need to
utilize progenitors with velocity dispersions as small as
$\sigma=32~{\rm km~s^{-1}}$.  These small halos, however, should be
excluded from contributing to the final mass by the large recoil
velocities. Another way of stating our result is to note that in our
fiducial model, but with $\sigma_{\rm min}=50, 100,$ or $200~{\rm
km~s^{-1}}$, corresponding to typical recoil velocities of $100, 200,$
or $400~{\rm km~s^{-1}}$, the final SMBH mass at $z=6.43$ is
$5.2\times10^8~{\rm M_\odot}$, $1.2\times10^7~{\rm M_\odot}$, and
$1.2\times10^5~{\rm M_\odot}$, respectively - an order of magnitude or
more below the inferred BH mass of SDSS 1054+1024.

In order to assess the sensitivity of the result above to our
assumptions, we here vary each of the parameters of our model.  For
each combination of $M_{\rm halo}$, $\eta$, $M_{\rm seed}$, and
$\epsilon$, we solve equation~\ref{eq:Mbh} with its left hand side set
to $M_{\rm bh}=4.6 \times 10^9\eta^{-1}~{\rm M_\odot}$, and the halo
mass set to $M_{\rm halo}$ at $z=6.43$, as discussed above.  We
further specify $M_{\rm seed}$ and $\epsilon$, and then find $v_{\rm
kick}$ by a Newton--Rhapson method.

We find that our results are insensitive to the adopted values of
$M_{\rm halo}$ and $\eta$.  Keeping all other parameters fixed at
their fiducial values, increasing or decreasing $M_{\rm halo}$ by a
factor of three yields $v_{\rm kick}=82.2~{\rm km~s^{-1}}$ and $v_{\rm
kick}=49.6~{\rm km~s^{-1}}$, respectively.  This is not surprising,
and reflects the fact that the rare, massive halos at the tail of the
mass function at $z=6.43$ have similar merging histories.  Similarly,
increasing or decreasing $\eta$ by a factor of three, we find $v_{\rm
kick}=80.4~{\rm km~s^{-1}}$ and $v_{\rm kick}=50.6~{\rm km~s^{-1}}$,
respectively.  The sensitivity to the value of the final BH mass is
only logarithmic because of the exponential growth predicted in
equation~\ref{eq:Mbh}.

The sensitivity to the adopted value of the seed mass, $M_{\rm seed}$,
is the same as to $\eta$ -- changing the mass of the seeds or of the
final BH is equivalent in our prescription,\footnote{Note that the
choice of $\eta$ and $\epsilon$ have a small additional effect through
the change they cause to the duty cycle, and hence to the inferred
value of the halo mass. For sake of clarity, we chose here to neglect
this effect, and parameterize the halo mass independently.} since the
model outlined above predicts the ratio $M_{\rm bh}/M_{\rm
seed}$. However, the seed mass is more uncertain than the inferred BH
mass.  Once again keeping all the other parameters fixed at their
fiducial values, we find that the choice of $M_{\rm seed}=1, 10^3$,
and $10^5~{\rm M_\odot}$ results in $v_{\rm kick}=50.6, 157.8$, and
$332~{\rm km~s^{-1}}$, respectively.  Indications from recent 3D
simulations (Abel, Bryan \& Norman 2000, 2002; Bromm, Coppi \& Larson
1999, 2002) are that the mass of the first, metal--poor stars are a
few $\times 10^{2} \, {\rm M_\odot}$. Nonrotating stars with masses
between $\sim 40-140~{\rm M_\odot}$ and above $\sim 260~{\rm M_\odot}$
collapse directly into a BH without an explosion, whereas stars in the
range $\sim 140-260~{\rm M_\odot}$ explode without leaving a remnant
(Heger et al. 2003).  Seeds as large as $10^5~{\rm M_\odot}$ could
only arise from the post--Newtonian instability of extremely massive
stars.

Finally, the largest sensitivity of our result is to the value of
$\epsilon$.  This is because $\epsilon$ directly enters the e--folding
time for the growth of the BH mass in equation~\ref{eq:Mbh}.  As an
example, we find that the choice of $\epsilon=0.05$, and $0.2$ result
in $438~{\rm km~s^{-1}}$ and $8.26~{\rm km~s^{-1}}$, respectively.
Thus, if the growth of each seed BH is as rapid as it would be with a
typical radiative efficiency of $\sim 0.05$, the typical recoil
velocities predicted by Favata et al. (2004) and Merritt et al. (2004)
would still allow the build--up of the SMBH in the $z=6.43$ quasar.
Note that in this case, most of the final black holes mass at $z=6.43$
would arise from the single most massive progenitor, starting to grow
at $z\approx 10$.  This is in contrast with our fiducial case with
$\epsilon=0.1$, which would imply that most of the BH mass was
assembled by the addition of $\sim 20$ seeds, each of which started to
grow at $z\sim 18$.  A typical efficiency of $\sim 0.05$ would be
significantly below the value obtained recently by Yu \& Tremaine
(2002).  Yu \& Tremaine (2002) compare the energy density in quasar
light and the mass density of local SMBHs (So\l tan 1981) as a
function of quasar luminosity. Interestingly, they find that bright
quasars (with luminosities similar to that of SDSS 1054+1024, or BH
masses of $\gsim 10^9~{\rm M_\odot}$) have a typical efficiency close
to $\epsilon \sim 0.2$.  This value is inferred only statistically,
and therefore represents the average efficiency of the entire
population of bright quasars at lower redshifts. Our results suggests
that such a high value cannot hold for the past history of the
individual source SDSS 1054+1025.

The constraints obtained here could potentially be avoided if BH seeds
at high redshift form only in a small fraction of the early dark
matter halos (Madau \& Quataert 2004).  In this case, when a dark
matter halo, carrying a seed BH, merges with another dark halo, the
seed BH would not experience a merger with another BH - it would then
not be ejected, and it could continue to grow unimpeded.  A small BH
``occupation fraction'' $f_{\rm bh}$ would make the predicted
abundance of high-$z$ quasars rarer by a factor of $f_{\rm bh}$, and
the discovery of the $z\sim 6$ SDSS quasars could then be challenging
to explain for $f_{\rm bh}\ll 1$. However, the fraction of DM halos
harboring SMBHs would be expected to increase over time, and $f_{\rm
bh}$ could increase significantly between $z\gsim 10$ and $z\sim 6$
(Menou et al. 2001). More detailed models, following the merger
history of BHs (Kauffmann \& Haehnelt 2000; Menou et al. 2001;
Volonteri et al. 2003; Islam et al. 2003), are needed to quantify this
scenario.

Merging SMBHs at $z\gsim 6$ can loose a non-negligible fraction of
their total mass to gravitational waves, especially if they typically
spin rapidly, and if they suffer a large number of mergers during
their assembly history (Menou \& Haiman 2004).  More generally,
although the mean expected recoil velocity is only weakly dependent on
the BH spin (Favata et al. 2004), a high value for the typical spin of
the merging BHs would strengthen our results through (i) the loss of
the BH mass to gravity waves, and (ii) through the larger expected
value of the radiative efficiency $\epsilon$, increasing the accretion
time-scale as $\epsilon^{-1}$.

\section{Conclusions}
\label{sec:conclude}

The recent discovery of luminous quasars at redshift $z>6$ provides
evidence that supermassive black holes (SMBH) as large as several
$\times10^9~{\rm M_\odot}$ were assembled during the first $\lsim
10\%$ of the current age of the universe.  In the context of
hierarchical structure formation scenarios, these early SMBHs grow via
accretion and mergers of seeds that appear at much earlier epochs,
$z\gsim 10$.  Unless the growth by accretion of individual seed BHs is
significantly faster than Eddington--limited accretion at a fiducial
radiative efficiency of $\epsilon\approx 10\%$, we find that such
scenarios appear inconsistent with the large recoil velocities,
$100~{\rm km~s^{-1}}\lsim v_{\rm kick} \lsim 600~{\rm km~s^{-1}}$,
that were recently calculated to occur during the coalescence of
massive BHs.

A natural resolution of this discrepancy would be for individual SMBHs
to grow in mass at significantly super--Eddington rates. In this case,
the SMBHs can arise from seeds that appear relatively more recently,
(i.e. closer to redshift $z\sim 6$, where the bright quasars exist),
in DM halos whose potential wells are deep enough to retain these
seeds despite their recoil.  Forthcoming instruments, such as the {\it
James Webb Space Telescope (JWST)} in the infrared, or deep radio
instruments, such as the {\em Allen Telescope Array (ATA)}, {\em
Extended Very Large Array (EVLA)}, and the {\em Square Kilometer Array
(SKA)}, whose sensitivities allow them to detect BHs as small as
$10^5~{\rm M_\odot}$ at $z\sim 10$ (see Haiman \& Quataert 2004 for a
recent review) would be able to shed light on whether or not the seeds
of $z\sim 6$ quasar black holes indeed extend out to these redshift.

\acknowledgements

The author thanks Kristen Menou and David Merritt for comments on an
earlier draft of this paper, and gratefully acknowledges financial
support from NSF grants AST-03-07200 and AST-03-07291.


\begin{references}

\reference{abn00} Abel, T., Bryan, G. L., \& Norman, M. L. 2000, ApJ, 540, 39

\reference{abn02} Abel, T., Bryan, G. L., \& Norman, M. L. 2002, Science, 295, 93

\reference{ar02}  Aller, M. C., \& Richstone, D. 2002, AJ, 124, 3035 

\reference{bl03a} Barkana, R., \& Loeb, A. 2003, Nature, 421, 341

\reference{b02} Begelman, M. C. 2002, ApJ, 568, L97

\reference{br61} Bonnor, W. B., \& Rotenberg, M. A. 1961, Proc. R. Soc. London, A265, 109

\reference{bcl01} Bromm, V., Coppi, P. S., \& Larson, R. B. 1999, ApJ, 527, 5 

\reference{bcl02} Bromm, V., Coppi, P. S., \& Larson, R. B. 2002, ApJ, 564, 23

\reference{cba84} Carr, B. J., Bond, J. R., Arnett, W. D. 1984, ApJ, 277, 445

\reference{eetal94} Elvis, M., et al. 1994, ApJS, 95, 1

\reference{fetal00} Fan, X., et al. 2000, AJ, 120, 1167

\reference{fetal01} Fan, X., et al. 2001, AJ, 122, 2833

\reference{fetal03} Fan, X., et al. 2003, AJ, 125, 1649

\reference{fhh04} Favata, M., Hughes, S. A., \& Holz, D. E., ApJL, submitted, astro-ph/0402056

\reference{f02} Ferrarese, L. 2002, ApJ, 578, 90

\reference{f83} Fitchett, M. J. 1983, MNRAS, 203, 1049

\reference{hc02} Haiman, Z., \& Cen, R.\ 2002, ApJ, 578, 702

\reference{hco04} Haiman, Z., Ciotti, L. \& Ostriker, J. P. 2004, ApJ, in press, astro-ph/0304129

\reference{hl01} Haiman, Z., \& Loeb, A. 2001, ApJ, 552, 459

\reference{hetal03} Heger, A., et al. 2003, ApJ, 591, 288

\reference{its03} Islam, R. R., Taylor, J. E., \& Silk, J.\ 2003, MNRAS, 340, 647

\reference{kh00} Kauffmann, G., \& Haehnelt, M.\ 2000, MNRAS, 311, 576

\reference{jetal01} Jenkins, A. et al. 2001, MNRAS, 321, 372

\reference{lc93} Lacey, C., \& Cole, S. 1993, MNRAS, 262, 627

\reference{madau04} Madau, P., Rees, M. J., Volonteri, M., Haardt, F., Oh, S. P. \& \ 2004, ApJ, 604, 484

\reference{mq04} Madau, P., \& Quataert, E. 2004, ApJL, submitted, astro-ph/0403295

\reference{mh04} Menou, K., \& Haiman, Z. 2004, in proceedings of "Black Hole Astrophysics 2004", Pohang, S. Korea, Journal of the Korean Physical Society (Special Issue), in press

\reference{mhn01} Menou, K., Haiman, Z., \& Narayanan, V. K.\ 2001, ApJ, 558, 535

\reference{metal04} Merritt, D., Milosavljevic, M., Favata, M., Hughes, S. A., \& Holz, D. E., ApJL, submitted, astro-ph/0402057


\reference{mh04} Mesinger, A., \& Haiman, Z. 2004, in preparation

\reference{p62} Peres, A. 1962Phys. Rev. 128, 2471

\reference{ps74} Press, W. H., \& Schechter, P. L. 1974, ApJ, 187, 425

\reference{st99} Sheth, R. K., \& Tormen, G. 1999, MNRAS, 308, 119 

\reference{s82} So\l tan, A. 1982, MNRAS, 200, 115 

\reference{setal03} Spergel, D. N. et al. 2003, ApJS, 148, 175

\reference{vhm03} Volonteri, M., Haardt, F., \& Madau, P.\ 2003, ApJ, 582, 559

\reference{wmj03} Willott, C. J., McLure, R. J., \& Jarvis, M. J. 2003, ApJ, 587, L15

\reference{yt02} Yu, Q. \& Tremaine, S. 2002, MNRAS, 335, 965

\end{references}
\end{document}